\documentclass[usenatbib]{mn2e}

\usepackage{astrojournals}
\usepackage{graphicx} 
\usepackage{amssymb,amsmath}
\usepackage{times}
\usepackage{url}

\newcommand{\beq}{\begin{equation}}
\newcommand{\eeq}{\end{equation}}

\newcommand{\kps}{\, {\rm km}/{\rm s}}

\newcommand{\Gyr}{\,{\rm Gyr}}

\newcommand{\kpc}{\, {\rm kpc}}

\newcommand{\LsunV}{{\rm L}_{\odot,{\rm V}}}

\providecommand{\ga}{\gtrsim}
\providecommand{\la}{\lesssim}

\begin{document}
\bibliographystyle{mn2e}

\title [Infall Times for MW Satellites]{Infall Times for
Milky Way Satellites From Their Present-Day Kinematics}

\author[Rocha et al.] 
{
Miguel Rocha,$^1$\thanks{E-mail: rocham@uci.edu} Annika
H. G. Peter,$^{1}$\thanks{E-mail: annika.peter@uci.edu} and James
Bullock,$^1$\thanks{E-mail: bullock@uci.edu}\\ $^1$Center for
Cosmology, Department of Physics and Astronomy, University of
California, Irvine, CA 92697-4575, USA\\
}

\maketitle
\date{\today}

\begin{abstract} 
We analyze subhalos in the Via Lactea II (VL2) cosmological simulation to look for correlations among their infall times and $z = 0$ dynamical properties. We find that the present- day orbital energy is tightly correlated with the time at which subhalos last crossed into the virial radius. This energy-infall correlation provides a means to infer infall times for Milky Way satellite galaxies. Assuming that the Milky Way's assembly can be modeled by VL2, we show that the infall times of some satellites are well constrained given only their Galactocentric positions and line-of-sight velocities. The constraints sharpen for satellites with proper motion measurements. We find that Carina, Ursa Minor, and Sculptor were all accreted early, more than 8 Gyr ago.  Five other dwarfs, including Sextans and Segue 1, are also probable early accreters, though with larger uncertainties. On the other extreme, Leo T is just falling into the Milky Way for the first time while Leo I fell in $\sim  2$ Gyr ago and is now climbing out of the Milky Way's potential after its first perigalacticon. The energies of several other dwarfs, including Fornax and Hercules, point to intermediate infall times, $2 - 8$ Gyr ago. We compare our infall time estimates to published star formation histories and find hints of a dichotomy between ultrafaint and classical dwarfs. The classical dwarfs appear to have quenched star formation after infall but the ultrafaint dwarfs tend to be quenched long before infall, at least for the cases in which our uncertainties allow us to discern differences. Our analysis suggests that the Large Magellanic Cloud crossed inside the Milky Way virial radius recently, within the last $\sim 4$ billion years.
 \end{abstract}


\section{Introduction}
The Milky Way is a unique laboratory for understanding the lives of dwarf galaxies ($L \lesssim 10^8M_\odot$).  Dwarf spheroidal galaxies, in particular, stand out among galaxies because of their high dark matter content, lack of gas, and lack of recent star formation.   Like larger galaxies \citep{dressler1980,butcher1984,goto2003}, dwarf galaxies appear to have a ``morphology--density'' relation, with dwarf spheroidal galaxies preferentially crowding around normal galaxies (or within groups) instead of  the field \citep{mateo1998,weisz2011}.  All of the galaxies within the Milky Way's dark-matter halo except the two Magellanic Clouds are dwarf spheroidals.  

In galaxy groups and clusters, the morphology--density relation is associated with a color--density relation or a star-formation--density relation \citep{kauffmann2004,blanton2005}.  The origin of these relations is thought to result from the quenching of star formation once galaxies become satellites in larger systems \citep[see, e.g.][]{berrier2009}.  Once inside the virial radius of a larger host, star formation in the satellites may be quenched either because they stop accreting fresh gas \citep[``strangulation'';][]{larson1980,bekki2002} or because their cool gas is stripped away \citep[``ram-pressure stripping'';][]{gunn1972} due to interactions with the host's gas halo.  High-speed encounters with other satellite galaxies or the host itself may similarly affect morphologies and star formation \citep[``harassment'';][]{moore1996}.  These processes are also relevant for dwarf galaxies around the Milky Way.  The Milky Way is likely surrounded by a hot gas halo of its own, which can aid in quenching star formation once galaxies fall within its reach 
\citep{maller2004,fang2006,peek2007,kaufmann2008,grcevich2009,grcevich2010}.  In addition, dwarfs can experience dynamical transformation due to tidal stirring \citep{lokas2010,kazantzidis2011}, though the smallest galaxies may very well be born puffy well before entering the halo \citep[][]{kaufmann2007}.

One way in which dwarf satellites are different than larger satellite galaxies in groups and clusters is that their low mass makes them inherently fragile, thus more susceptible to quenching processes that would not otherwise affect $\sim L_*$ galaxies.
Specifically, it is possible that some of the known dwarf spheroidals were quenched prior to infall into the Milky Way.   During and after reionization, photons pour into the intergalactic medium, heating and pressurizing the gas so much that it is unable to collapse onto dark-matter halos with circular velocity smaller than $V_{max} \sim 30\hbox{ km s}^{-1}$ 
\citep[the exact value of which is disputed;][]{thoul1996,bullock2000,benson2002,benson2003,dijkstra2004,okamoto2008}.  Reionization photons may also photoevaporate gas already present in halos before reionization \citep{barkana1999}.  Dwarf-galaxy dark-matter halos have small escape velocities; therefore, stellar winds or supernovae may permanently blow gas out of these small galaxies \citep{governato2010}.  The newly discovered population of ultrafaint dwarf spheroidals in the Milky Way \citep{willman2005,belokurov2007,kirby2008} have overwhelmingly old stellar populations and are often speculated to be ``fossils of reionization''---galaxies that only form stars prior to reionization  \citep[e.g.,][]{martin2008b,madau2009,bovill2010a,bovill2010b}.  However, it remains unclear whether the ultrafaints are old because they are true fossils of reionization or simply because they fell into the Milky Way at an early epoch.

One way we can hope to discriminate quenching scenarios is to determine when each galaxy became a satellite and then compare this to an inferred star formation history.    The Milky Way is a unique laboratory for answering these questions not only because it is currently the only place where we can find ultrafaint galaxies but because of the availability of exquisite photometric and kinematic data for virtually all of the satellites.  We have three-dimensional configuration-space positions and line-of-sight velocities for every dwarf satellite (the dwarfs considered here are listed in Table 1).  Moreover, for a subset of the classical dwarf galaxies, we also have proper motions.  Thus, we have four- or six-dimensional phase space positions for the Milky Way dwarf satellite galaxies.  These kinematic data allow us to estimate the infall times of the dwarfs.

Previous work has focused on constraining the dwarf infall times by evolving satellite orbits back in time based on those observed phase-space coordinates today or by tracing specific satellite orbits forward in time in large N-body simulations \citep{besla2007,lux2010,angus2011,boylan-kolchin2011}.  The problem with these approaches is that they are sensitive to Poisson noise---specific things like the choice of the triaxiality of the Milky Way and its evolution through time, satellite interactions in the simulated Milky Way, treatment of dynamical friction and tidal stripping of the satellites, all cause large uncertainties for the infall times of individual orbits.  

Instead, we adopt a simpler, more statistical approach to determining the infall times of the Milky Way dwarf satellites (including the dwarf irregular Magellanic Clouds).  In particular, we focus on using a simulation of a Milky-Way-type halo to determine an infall-time probability distribution function (PDF) for each dwarf based on simulated subhalos with similar present-day phase-space coordinates.  In Sec. \ref{sec:methods}, we describe the properties of the simulation that are relevant to this work.  We use the simulation to show, in Sec. \ref{sec:einfall}, that there is a strong correlation between the infall time of subhalos that might host dwarf galaxies to their binding energy today to the host dark-matter halo.  We call this correlation the ``energy-infall'' relation.  Since the simulated halo is similar to the halo that hosts the Milky Way, we make the ansatz that the energy-infall relation of Sec. \ref{sec:einfall} can be applied to Milky Way dwarf galaxies.  In Sec. \ref{sec:accMW}, we create PDFs of the dwarf infall times based on the subhalos that have galactocentric positions, line-of-sight velocities, and proper motions (if measured) within the measurement error bars of observed dwarf galaxies.  We can constrain the infall times using the energy-infall relation because the kinematic data yield estimates of the binding energy (or upper limits thereof if no proper motion measurements exist).  In Sec. \ref{sec:discussion}, we show how the infall-time PDFs correspond to existing determinations of the star-formation histories of the Milky Way's dwarfs.  We discuss what the implications are for quenching mechanisms for dwarf spheroidals and trends between galaxy properties and environment.  In addition, we highlight further applications of the energy-infall relations in the study of galaxy evolution.  We summarize our conclusions in Sec. \ref{sec:conclusion}.  

\section{Methods}\label{sec:methods}

We use the Via Lactea II (VL2) simulation for our analysis of satellite orbits and infall times \citep{diemandetal2007,diemandetal2008,kuhlen2010}.  The VL2 simulation is a high-resolution lambda cold dark matter ($\Lambda$CDM) cosmological simulation that focuses on a dark matter halo of approximately the same size as the one that hosts the Milky Way, with a maximum circular velocity $V_{max} = 201~\hbox{km/s}$ at $z=0$.  

The cosmology assumed in this simulation is taken from the flat-universe six-parameter analysis of the \emph{WMAP} three-year data set \citep{spergel2007}: $\Omega_m = 0.238$, $\Omega_\Lambda = 0.762$, $h=0.73$, $n_s = 0.951$, $\sigma_8 = 0.74$.  The resolution of VL2 is high enough, with particle masses of $4.1\times 10^3 M_\odot$ each, to
resolve thousands of subhalos bound to the main host. We have selected a
sample of $\sim2000$ bound subhalos with maximum circular velocities $V_{max} > 5 \hbox{ km/s}$ at $z=0$ and studied their dynamics at redshift $z=0$. Halos in VL2 are found through the 6DFOF halo finder described in \citet{diemand2006}. The $V_{max}$ threshold guarantees that we do not miss any subhalos that might plausibly be hosting dwarf galaxies in the Milky Way today.

Throughout this work, we define the main-halo mass $M_{200}(z)$ and virial radius $R_{200}(z)$ in terms of an overdensity $\Delta = 200\rho_m(z)$, where $\rho_m(z)$ is the homogeneous matter density in the Universe, and $M_{200} = 4\pi \Delta R_{200}^3/3$.  With this definition, the $z=0$ virial mass of the VL2 host is $M_{200} = 1.9\times 10^{12}M_\odot$ today, with a virial radius of just over 400 kpc.  We define the center of the halo to be the center of mass of bound particles within the virial radius.

The infall time of a subhalo into the main host depends somewhat on definition.  We identify $t_{infall}$ as the look-back time since the subhalo {\em last} crossed inward through the virial radius of the main halo $R_{200}(z)$.  We explored an alternative definition where $t_{infall}$ is identified with first crossing through the virial radius but found that the resulting kinematic correlations were 
more scattered (but still present), so we adopted the last-crossing definition.

%  Namely, it is more closely associated with the time at which the subhalo becomes bound to and must remain inside of the virial region of the halo.  Moreover, if we 
%compare the infall time as we have defined it to the star-formation histories of Milky Way dwarf galaxies, we can tell if star formation persisted even after the galaxies
 %were swallowed by the Milky Way halo. 

We define the binding energy of a subhalo to the host halo by
\begin{eqnarray}
  \mathcal{E} = -\phi(r) - \frac{1}{2}V^2,
\end{eqnarray}
where the gravitational potential 
\begin{eqnarray}
  \phi(r) = -\int_r^{R_{0}} \frac{GM(<r)}{r^2}
\end{eqnarray}
is defined such that $\phi(R_{0}) = 0$ for a reference radial position $R_{0}$.  The enclosed mass at galactocentric distance $r$ is $M(<r)$ and the subhalo velocity with respect to the halo center is $\mathbf{V}$.  For this work, we define $R_{0} = 1$ Mpc in physical units so that the gravitational potential has a fixed zero point across cosmic time.  By contrast, if we had chosen $R_{0} = R_{200}$, the energy of a particle would change throughout time even if its orbit were fixed and the density profile were constant in time since $R_{200}$ grows with time.  

\section{The Energy-Infall Correlation}\label{sec:einfall}
We checked for correlations among subhalo infall times and many different subhalo orbital properties (orbit circularity, angular momentum, binding energy, radial velocity, current position, etc.).  Many of these properties showed no strong correlation with infall time, but there is a clear correlation with binding energy.
A key result of this paper is that the current binding energy of a subhalo $\mathcal{E}$ is a simple, clean $z=0$ indicator of the subhalo's infall time. 

\begin{figure} \begin {center} %\resizebox{15cm}{12cm}{ 
\includegraphics[width=0.5\textwidth]{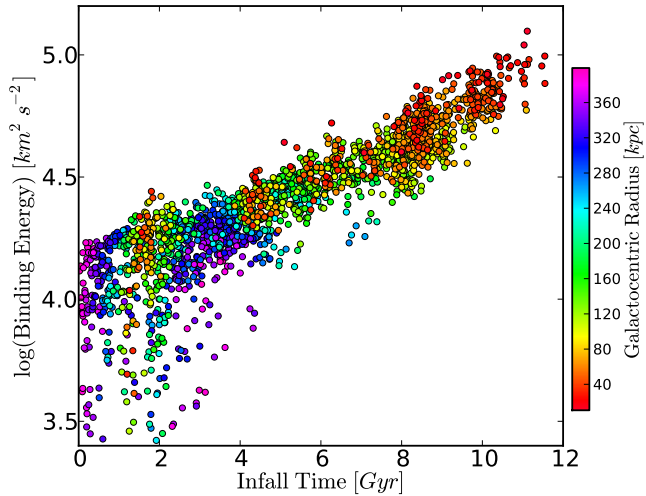}%}
\end {center} 
\caption{Binding energy vs. infall time for the selected sample
of VL2 subhalos at $z=0$. Colors indicate galactocentric distance. Notice how the least
bound subhalos are the ones accreted most recently and also the only ones with
large galactocentric distances.}
\label{TvsE.fig} 
\end{figure}

The energy-infall correlation is demonstrated in Fig. \ref{TvsE.fig}, in which we also color-code the subhalos by their current radial position with respect to the halo center.  We find that subhalos accreted early are more tightly bound to the halo with relatively little scatter, and that late-infall subhalos have lower binding energy (albeit with more scatter).  Another way of looking at this correlation is that we find that subhalos that are currently deep within the potential well of the halo preferentially have higher binding energy and earlier infall times (greater $t_{infall}$) than subhalos that orbit farther out.  While this is generally expected from hierarchical structure formation theory, the quantification of this relation we find in this work proves highly useful, and we return to this point in later sections.

\begin{figure} \begin {center} %\resizebox{15cm}{12cm}{ 
\includegraphics[width=0.5\textwidth]{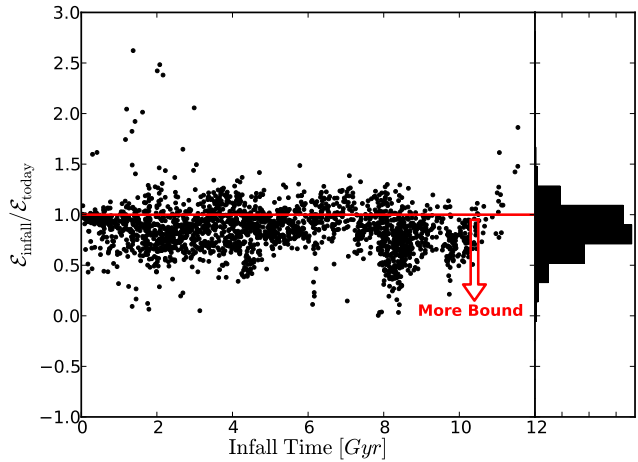}%}
\end {center} 
\caption{Ratio of the binding energy at infall to the binding energy today
as a function of subhalo infall times. On average the subhalos in the sample are slightly more bound today than when they fell into the VL2 halo, but clearly the binding energy of subhalos today tracks the binding energy at infall regardless of the infall time.}
\label{TvsEratio.fig} 
\end{figure}

We can trace this correlation directly to the energy of the subhalos at their infall epochs.  In Fig. \ref{TvsEratio.fig}, we show the ratio of the binding energy of subhalos today to the binding energy at $t_{infall}$, as a function of $t_{infall}$.  Recall that $\mathcal{E}$ is larger for more tightly bound subhalos, so that the $\mathcal{E}_{infall}/\mathcal{E}_{today} < 1$ for halos that become more bound after infall and $\mathcal{E}_{infall}/\mathcal{E}_{today}>1$ if the subhalo becomes less tightly bound towards $z=0$.  This figure demonstrates that while there is some scatter about $\mathcal{E}_{infall}/\mathcal{E}_{today} = 1$, the average value of this ratio as a function of infall time is nearly $1$ and does not change appreciably with infall time.  Thus, the subhalo on average conserve their energies at infall, with $\mathcal{E}_{infall} \sim -\phi(R^{infall}_{200}) - \frac{1}{2}(V^{infall}_{200})^2$.   Because the binding energy at infall is linked to the virial properties of the host halo at that time, the correlation of binding energy with infall time arises from the mass-assembly history of the halo.  

Why are subhalo energies, on average, conserved throughout cosmic time?  For the moment, we only consider changes of energy due to interactions between subhalos and the host.  Three-body interactions including another subhalo may also serve to increase subhalo energies, or reduce binding energies 
\citep{sales2007,donghia2008,ludlow2009}, but we consider only those processes that cause subhalos to become more bound.  The energy may change if dynamical friction is significant or if the gravitational potential evolves.  Let us consider the case of dynamical friction.  Dynamical friction is only likely to change the energies of the most massive subhalos, since the merger timescale
\begin{eqnarray}
  \tau_{merge} \propto (M_{200}/M_{sat})^{1.3} t_{dyn},
\end{eqnarray}
where $t_{dyn}$ is the typical dynamical time in the host galaxy (mass $M_{200}$), and the satellite mass is $M_{sat}$ \citep{boylan-kolchin2008}.  The merger timescale is linked to the timescale over which the energy changes, since it takes a time $\tau_{merge}$ for the satellite to go from an initial binding energy $\mathcal{E}_{infall}$ to $\mathcal{E}_{today} = -\phi(0)$.  For a Milky Way-mass halo, the dynamical time is roughly 1 Gyr, and so only the subhalos that have $M_{sat}/M_{200} \gtrsim 0.1$ will have merged with the host halo in a Hubble time.  Thus for most $z=0$ subhalos, dynamical friction will have only a modest effect on the binding energy, in the direction of making them increasingly more bound.  Those subhalos for which dynamical friction is important in changing the energy significantly are also those that are most likely to have already merged with the host or have been tidally shredded, and are thus not part of the surviving $z=0$ subhalo population. 
%JB - I think its best if we leave this last bit out.  The MW has SMC, LMC, and Sag, which all likely have some degree of dynamical friction that has affected them.  What you've written stands for the bulk of the population, which sets the trend.   
%-->
%In addition, the Milky Way is unlikely to have suffered 10:1 mergers in the recent past given the thinness of the thin disk \citep{purcell2009}.  Therefore, we do not expect to see satellites in the Milky Way today for which dynamical friction has acted in a significant way.

Let us consider the second case of changes to the halo potential.  Dark-matter halos typically have a ``fast'' growth stage, within which the matter within the scale radius of the density profile today is rapidly acquired, and a ``slow'' growth stage, after which the halo grows constantly and without major mergers \citep{wechsler2002,zhao2009}.  These two regimes are artifacts of the shape of the linear density perturbation, with the fast growth regime linked to regions in which the density perturbation $\delta \sim \hbox{const}$, and the slow growth regime to regions in which the $\delta$ is a sharply falling function of distance from the center of the perturbation \citep{dalal2010}.  For most of the VL2 halo's history, it is in the slow growth period, meaning that the density profile near and outside the virial radius is a sharply falling function of distance.  If we approximate the local density profile as a power law, then $\rho \propto r^{-\alpha}$, the enclosed mass $M(<r) \propto r^{3-\alpha}$ and the gravitational potential
\begin{eqnarray}
  \phi(r) &=& -\int_r^{R_{0}} \frac{GM(<r)}{r^2} \\
          &\sim& -\int_r^{R_{0}}  r^{1-\alpha} \\
          &\sim& R_{0}^{2-\alpha} - r^{2-\alpha}.
\end{eqnarray}
 Thus if the density profile falls off as $\alpha > 2$, which is certainly the case in the slow-growth phase, then $\phi(r) \propto r^{2-\alpha}$.  Since halos form from the inside out, the enclosed mass at a fixed radius within the virial radius does not change much with time \citep{book2011}.  As such, we do not expect the potential to change much with time at a fixed radius.

In summary, we find a strong correlation between the energy of subhalos at $z=0$ in the VL2 simulation and their infall times.  This is because the subhalos surviving to $z=0$ on average conserve their binding energies since their infall epoch, and all halos falling in at the same time must have roughly similar energies, of order the energy of a circular orbit at the virial radius.

\section{Infall Times of Milky Way Satellites}\label{sec:accMW}

Under the assumption that Milky Way satellite galaxies are hosted by dark
matter subhalos similar to those predicted by the VL2 simulation, and that the Milky Way itself and its mass-assembly history is something like the VL2 halo, we can estimate the infall times of Milky Way satellites based on their energies, which rely fundamentally on kinematic measurements.  The goal of this section is to show how the measured kinematics of the Milky Way dwarfs define a probability distribution function (PDF) for the infall time of each dwarf.  Ideally, we would use positions, line-of-sight velocities, and proper motions for each dwarf so that we could reconstruct each dwarf's orbital energy assuming that the gravitational potential of the Milky Way matches that of VL2.  We would not expect a delta-function like peak in the PDF for $t_{infall}$ even in this idealized case due to the scatter in the energy-infall relation, but we would expect a well-defined peak in the PDF (see Fig. \ref{TvsE.fig}).  From Fig. \ref{TvsE.fig}, we see that the position information should reduce the width of the PDF, especially for recently accreted subhalos. 

Unfortunately, for most Milky Way dwarf galaxies, especially the subset of ultrafaint dwarfs, we only have position measurements and line-of-sight velocities.  Thus, we begin by exploring how well one may constrain the infall time with only these four-dimensional (position vector and line-of-sight velocity) data.

\subsection{Constraints from radial velocities and distances alone}\label{sec:acc4d}

\begin{figure*} \begin {center} %\resizebox{15cm}{12cm}{ 
\includegraphics[width=\textwidth]{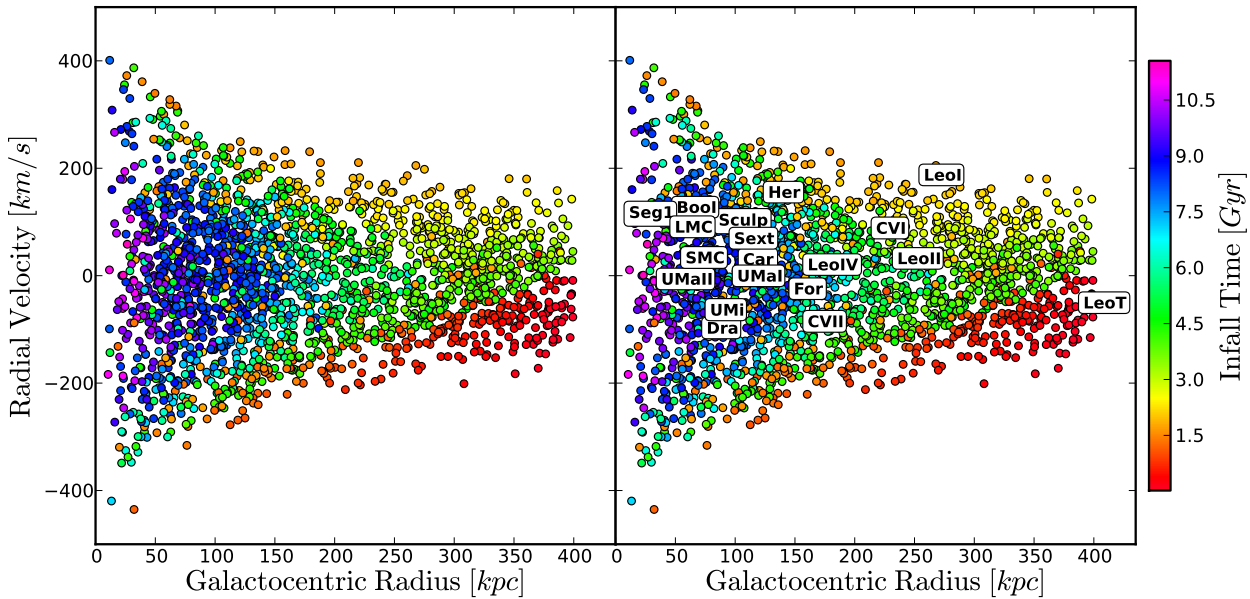}%}
\end {center} 
\caption{
Infall times in color as a function of the radial velocity and the
galactocetric distance of the VL2 sample of subhalos. The left panel shows
only the VL2 subhalos, the right panel shows a sample of Milky Way dwarf
galaxies on top of the VL2 subhalos. It is evident that with just the radial
velocity and galactocentric distance information coupled with the VL2
predictions we can set constraints on the infall times of Milky
Way dwarfs from their position on this space. It is specially clear that LeoT
and LeoI seem to have been accreted to  the Milky Way halo recently as 
opposed to Segue 1, which appears to be accreted early.}
\label{RvsVr.fig} 
\end{figure*}

In Fig. \ref{RvsVr.fig}, we present our subhalo population in the space of Galactocentric distance $r$ and radial velocity $V_r$.  The subhalo points are color coded by infall time, as indicated by the legend on the right.  The outer envelope in $r-V_r$ space is dominated by recently accreted subhalos; as one moves to smaller $|V_r|$ and especially as one moves to smaller $r$, the subhalos are accreted further back in time. The red points ($t_{infall} \sim 1$ Gyr) with negative velocities and large radii correspond to systems that are just falling in for the first time and the yellow points ($t_{infall} \sim 3$ Gyr) are systems that are just coming back out after their first pericenter passage.  The purple and blue points correspond to early accretions ($t_{infall} \sim 8-10$ Gyr) and naturally cluster at the small radii and fairly low speeds indicative of higher binding energy.

The right-hand panel of Fig. \ref{RvsVr.fig} is identical to the left except that we have overlayed the Galactocentric distances and line-of-sight velocities of Milky Way dwarfs.  The photometric and kinematic data on the Milky Way dwarf galaxies we consider are presented in Table 1.  We identify the measured line-of-sight velocity for the dwarfs with the radial velocity since the Sun is deep in the potential well of the Milky Way and the dwarfs are much farther out.    We can estimate the infall times
of dwarfs by comparing their positions on this diagram to the infall times of the VL2 subhalos that occupy similar positions.
For example, we can discern that the Leo T and Leo I dwarf spheroidals were accreted recently, with
$t_{infall} \lesssim 1 \hbox{ Gyr}$ and $t_{infall} \sim 2 \hbox{ Gyr}$ respectively.  We note that an infall time of $\sim 2$ Gyr for Leo I is consistent with the model reported by \cite{mateo2008} that gives the last pericenter crossing for Leo I as $\sim 1$ Gyr.   Conversely, 
Segue 1 appears to be an old companion of the Milky Way with $t_{infall}
\sim 8-10 \hbox{ Gyr}$. The infall times for other MW satellites are not as easily discerned by eye in this diagram, but many still provide useful constraints on the infall time even in the absence of 3d velocity data, as we will now discuss.

The solid black histograms in Fig. \ref{Tdists.fig} show infall-time PDFs for most of the known MW satellites.  We constructed these based on each galaxy's $r$ and $V_r$ values compared to those of the VL2 subhalos.  
The top three rows show the classical dwarf galaxies and the bottom four show the more newly discovered ultrafaint dwarf population.
The red histograms include proper motion information and will be discussed in Sec. \ref{sec:acc6d}.

The solid black PDFs for each dwarf $d$ are constructed by including the infall times for all VL2 subhalos with radii $r$ that obey $|r - r^d| \le  c\ \sigma_{r}$ and radial velocities $V_r$ that obey $|V_r - V_r^d| \le c \ \sigma_{V_r}$.  Here  $\sigma_{r}$ and $\sigma_{Vr}$
are the observational errors on radial velocity and galactocentric distance for 
the appropriate galaxy $d$.  The variable $c \ge 1$ is a proportionality constant set by the requirement
that there are at least 20 VL2 subhaloes in the subsample of each galaxy.  The positions, radial velocities, and uncertainties for the Milky Way satellites are summarized in Table \ref{accTimes.tab}.  

Based on their radial positions and velocities alone several of the dwarfs have reasonably well constrained infall times.  As previously discussed, Leo I and Leo T have been quite recently accreted.  Draco, Sextans, and Bo{\" o}tes likely fell into the Milky Way well before $z=1$ (8 Gyr), although the precise redshift beyond $z=1$ is not clear.  The infall times for other dwarfs are ambiguous.  In particular, Canis Venatici II and Willman I could have been accreted anytime in the past 10 Gyr.

\subsection{Including proper motions}\label{sec:acc6d}

Six MW dwarfs have published proper motion estimates from the \textit{Hubble Space Telescope} (\emph{HST}) with reasonably small errors: Ursa Minor, Carina, Sculptor, Fornax,  the Small Magellanic Cloud (SMC), and the Large Magellanic Cloud (LMC; see Table 1, which provides associated references in each case).  Leo II also has a published proper motion, but its relative error is fairly large \citep{lepine2011}.
For the galaxies with small errors we should be able to
to place better constraints on their accretion times as we may estimate directly the binding energy and not just place an upper limit on it.  

To demonstrate the improvement to the infall time constraints
coming from proper motions we examine a subsample of VL2 subhalos with 
similar radial velocities and galctocentric distances to each of the
MW dwarfs under study. Again, the subhalo sample for each Milky Way satellite is selected such that there are at least 20 subhalos in each sample according to the position and radial velocity constraints discussed in the previous section.  
Figure \ref{TvsVt.fig} shows the tangential velocities and
infall times of those subhalos associated with each of those
satellite galaxies for which proper motions are measured with high precision.  The grey band shows
the $1\sigma$ measurement of the proper motion for each dwarf.
It is clear from Fig. \ref{TvsVt.fig} that the extra information added from
proper motions can sharpen the infall-time PDFs.  

\begin{figure*} \begin {center} %\resizebox{15cm}{12cm}{ 
\includegraphics[width=\textwidth]{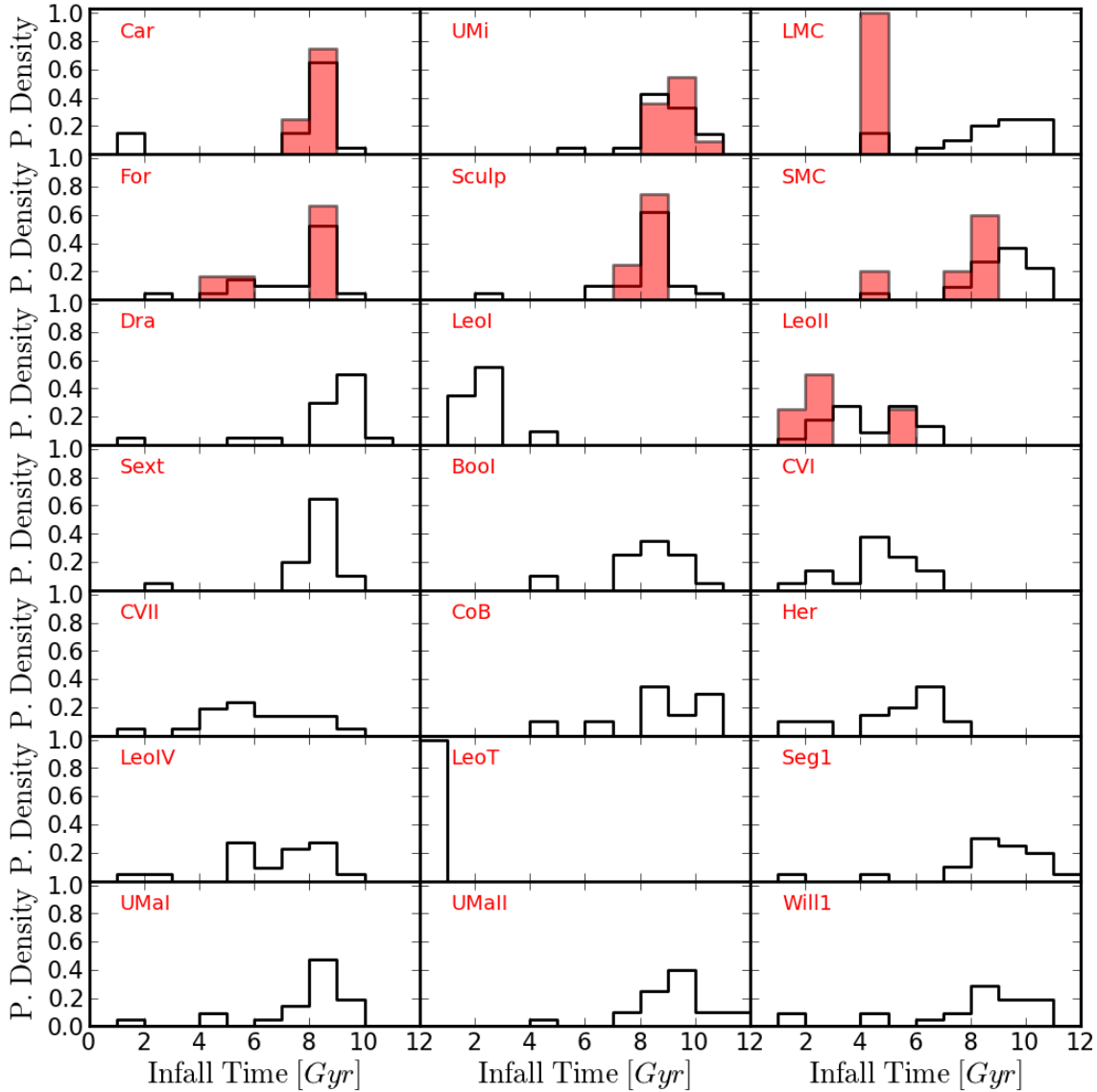}%}
\end {center} 
\caption{Infall time histograms for each dwarf constructed by matching kinematic properties of subhalos in VL2 to measured properties in the dwarfs (see text for details).
The solid black lines include only radial velocity and position information for the dwarfs.  The red histograms add the additional constraint associated with
proper motions when they are available.  When proper motions errors are relatively small (top two rows) this additional information tightens the inferred range of infall times significantly.}
\label{Tdists.fig} 
\end{figure*}

\begin{figure*} \begin {center} %\resizebox{15cm}{12cm}{ 
\includegraphics[width=\textwidth]{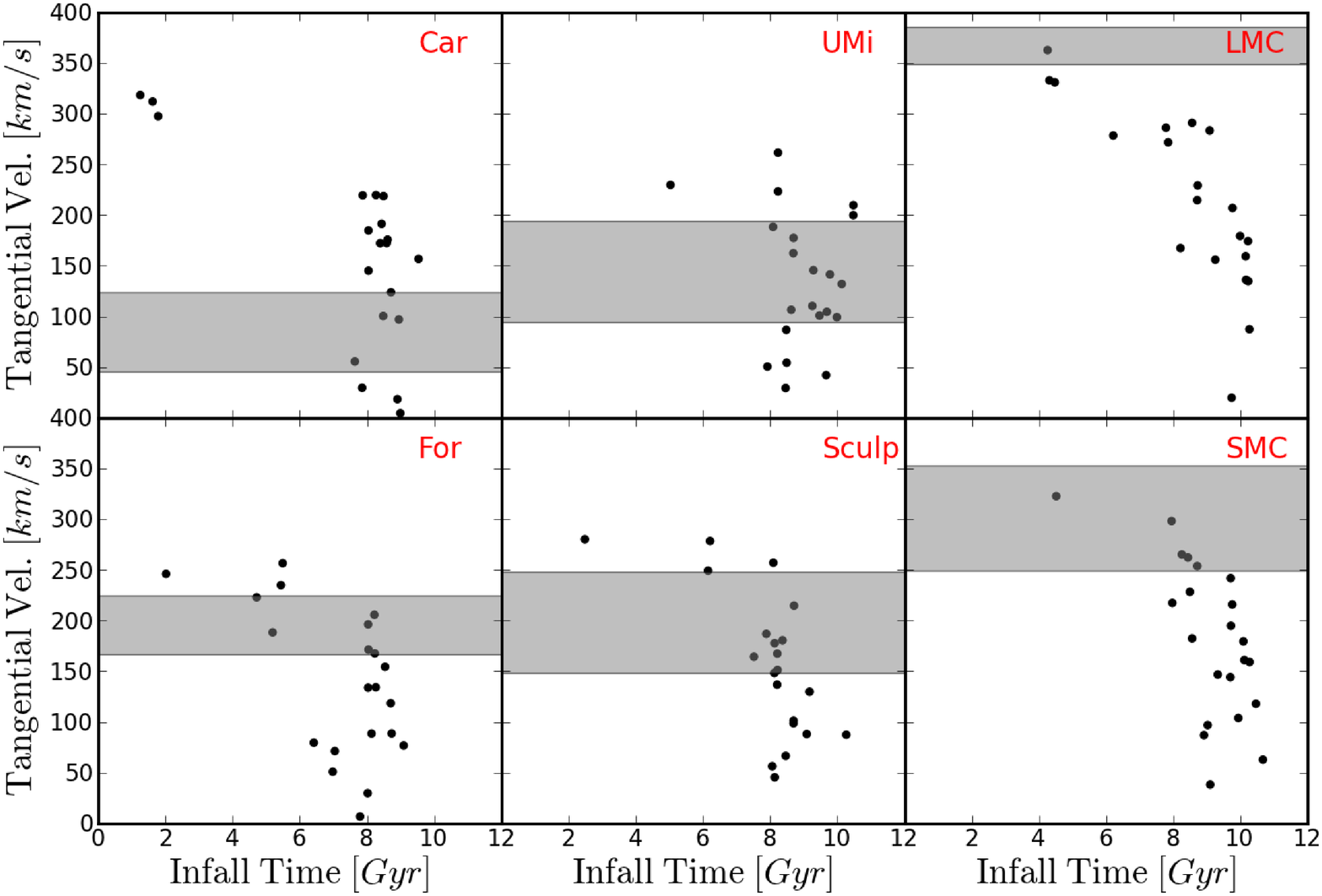}%}
\end {center} 
\caption{Tangential velocity as a function of infall time for subsamples of
subhalos with similar radial velocities and galactocentric distances to those of
the given dwarf galaxies. The subsample selection criterion is the same as in 
Fig. \ref{Tdists.fig}. The $1\sigma$ uncertainties in the proper motions are represented by the shaded regions. The addition of proper motion constraints provides a better estimate of the infall time than radial velocity alone.}
\label{TvsVt.fig} 
\end{figure*}

The red histograms
in the upper panels of Fig. \ref{Tdists.fig} demonstrate that the extra information does indeed sharpen the infall-time constraint relative to the constraint without proper motions.  For Carina, Ursa Minor, and Sculptor, their relatively small tangential velocities disfavor the tail of recent infall-times that would be allowed by their radial velocities alone (compare red to black solid histograms), pushing their inferred infall times towards earlier accretion ($t_{infall} \gtrsim 8$ Gyr).  
For the SMC, on the other hand, its relatively high proper motion favors later infall solutions, giving $t_{infall} \lesssim 4-9$ Gyr.  Interesting, these late infall solutions for the SMC were otherwise disfavored based on its radial velocity alone.  A similar story follows for the LMC, which must be accreted recently according to our comparison with VL2 subhalos ($t_{infall} \lesssim 4$ Gyr).  The measured tangential speed for Fornax is low enough to disfavor the possibility of very recent accretion, giving $t_{infall} \sim 5-8$ Gyr.  

Finally, the proper motions reported for Leo II provide little additional constraint on its infall time, though reassuringly  Leo II proper motions are consistent with those expected for subhalos of the appropriate distance and radial velocity.  It should be emphasized that the same agreement applies to the SMC and the LMC.  We do find VL2 subhalos in our sample that can be matched with their speeds and positions.
 
 Table 1 provides a summary of our results on infall times in the right-most column.
 
\subsection{Future observations}\label{sec:future}

\begin{figure*} \begin{center}
\includegraphics[width=\textwidth]{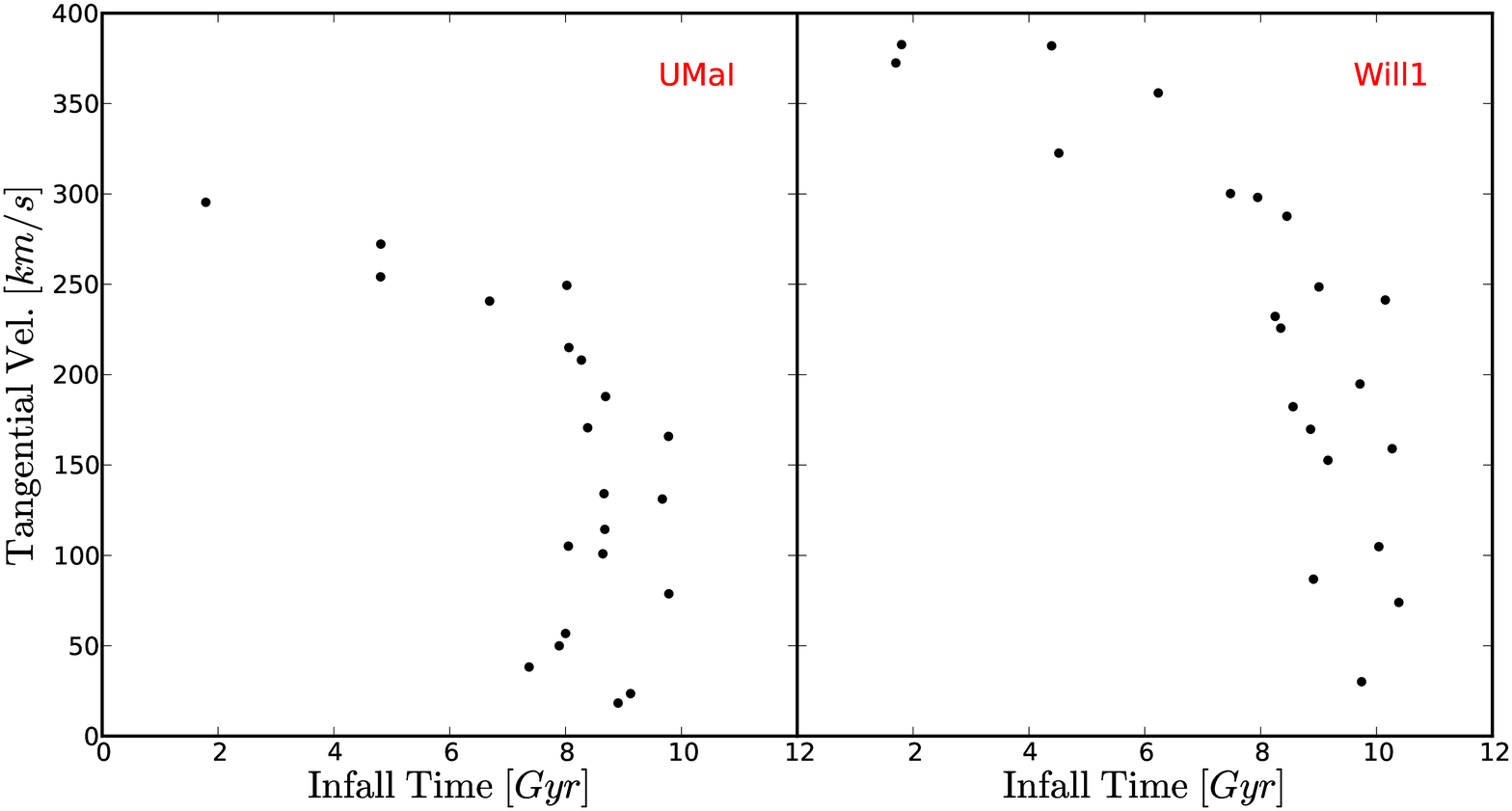}
\end{center}
\caption{Tangential velocity as a function of $t_{infall}$ for VL2 subhalos that have similar positions and radial velocities of Ursa Major I (\emph{left}) and Willman I (\emph{right}).  Measured proper motions with an uncertainty of order $\sim 50\hbox{ km s}^{-1}$ can bracket infall times to within $\sim 2$ Gyr in most cases.}
\label{VtUMaIWillI.fig}
\end{figure*}

Since a number of the dwarfs without proper-motion measurements at the present, especially the ultrafaint population, have broad infall-time PDFs, we would like to know how well one would need to measure the proper motions to get unambiguous infall times.  In Fig. \ref{VtUMaIWillI.fig}, we show scatter plots for the tangential velocity as a function of infall time for the VL2 subhalo samples associated with Ursa Major I and Willman I.  The nature of the latter object is under debate; it may be a disrupting star cluster \cite{willman2010}.  However, the distribution of tangential velocities as a function of infall time for associated subhalos is illustrative.  For both of these objects, if the tangential velocity is small ($\lesssim 200 \hbox{ km s}^{-1}$), a proper-motion measurement with an associated uncertainty in the tangential velocity of $\sim 50\hbox{ km s}^{-1}$ (typical of the uncertainty in the tangential velocity of the classical dwarfs) would be sufficient to bracket the infall time to within 2 Gyr.  The scatter in $t_{infall}$ for fixed tangential velocity is higher for large tangential velocities, but even an uncertainty of $50 \hbox{ km s}^{-1}$ would be sufficient to tell if these objects fell in early ($t_{infall} \gtrsim 8$ Gyr) or late.

\section{Discussion}\label{sec:discussion}
Here, we discuss the energy-infall relation and infall-time PDFs in several contexts.  First, we compare the infall-time PDFs for individual dwarf galaxies with their  star-formation histories.  Second, we speculate as to what kinds of theoretical and observational work would be required to produce more robust infall-time PDFs.  Third, we speculate on some possible applications of the energy-infall relation to tidal streams.

\subsection{Dwarf galaxy infall times and their star-formation histories}
One of the main reasons for determining the infall times of the Milky Way dwarf galaxies is to use that time with respect to the star-formation history in the dwarfs as a diagnostic for star-formation quenching mechanisms in these small galaxies. 
%In this section, we take the infall-time PDFs found in the previous section seriously and compare the infall times with the dwarf star-formation histories.  
Although the infall-time PDFs are based on a single simulation of a Milky-Way-like halo, we show that the infall-time PDFs have the power to show a number of interesting trends regarding quenching in small galaxies.  For this discussion, it will be helpful to the reader to have Fig. \ref{Tdists.fig} in view or to refer to the last
column in Table 1, where we have compiled our infall time estimates.

\subsubsection{Classical dwarfs}

Among the classical dwarf spheroidal galaxies, we find that Carina, Draco, Ursa Minor (UMi), and Sculptor likely fell into the Milky Way early ($t_{infall} > 8$ Gyr).   Of those four, UMi and Sculptor have old stellar populations, with no evidence of stars younger than $t \sim 10$ Gyr in UMi and $t\sim 7$ Gyr in Sculptor \citep{hurley1999,monkiewicz1999,mighell1999,wyse1999,carrera2002,dolphin2002,tolstoy2004,cohen2010,deboer2011}.  Draco likely has a predominantly old stellar population as well, and although the exact age of this is debated, most authors suggest most stars are $>10$ Gyr old \citep{grillmair1998,aparicio2001,orban2008}.  
The Carina dwarf is somewhat different.  It likely formed most of its stars in bursts at 3, 7 and 11 Gyr lookback time, with $\sim 50\%$ of the stars $\sim 7$ Gyr ago \citep{smecker1994,smecker1996,hurley1998,monelli2003,rizzi2003}---after the time we estimate that it fell into the Milky Way. 

\begin{figure*}
  \begin {center} %\resizebox{15cm}{12cm}{ 
\includegraphics[width=\textwidth]{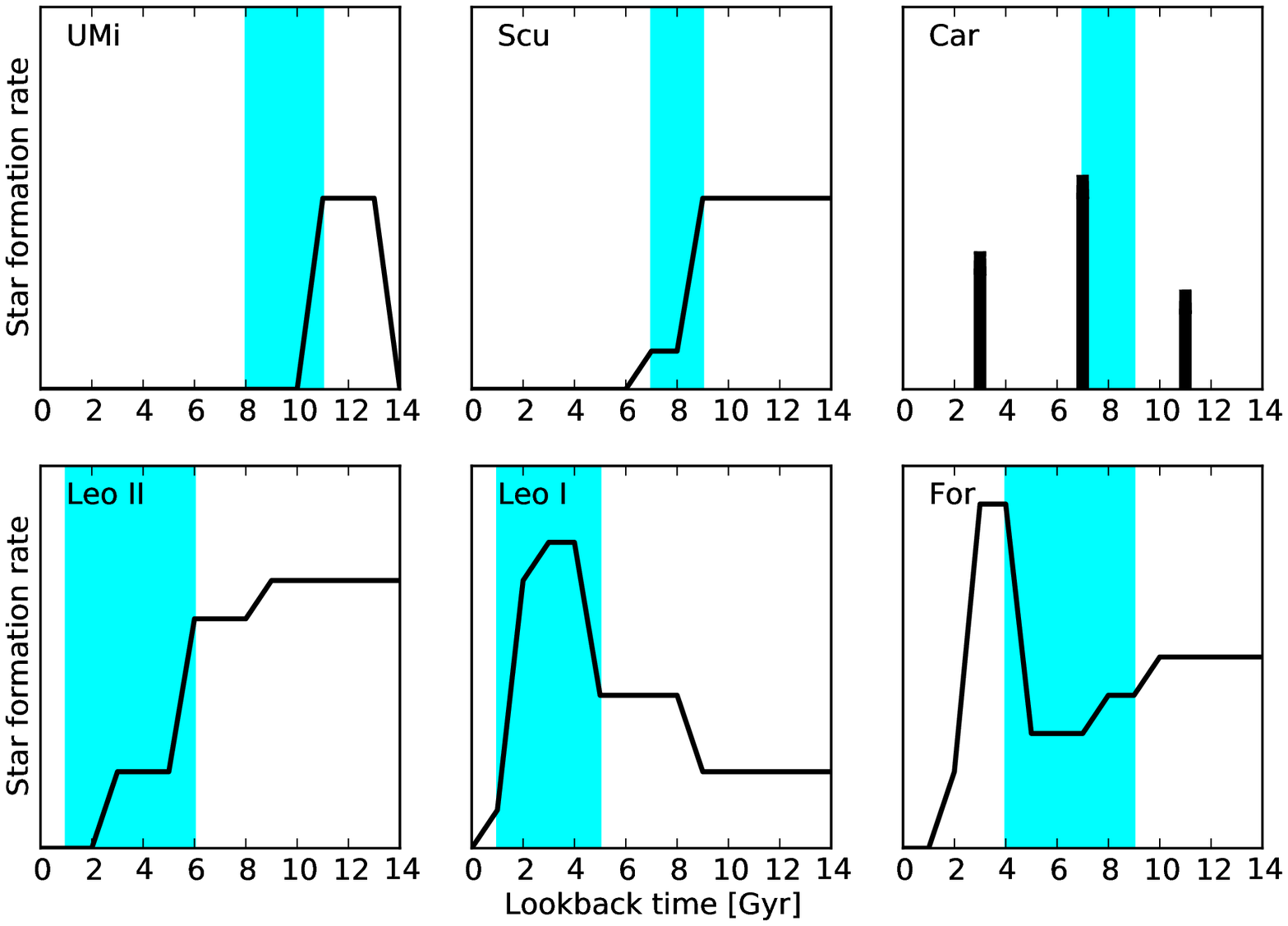}%}
\end {center} 
\caption{Qualitative depictions of star-formation histories (black solid lines) compared to our estimated infall times (cyan bands) for a sample of classical dwarf galaxies. The line heights in the Carina plot indicate the relative strengths of the star bursts.  The likely infall times are denoted by the shaded regions. The star-formation histories come from \citet{hurley1998} for Carina; \citet{dolphin2002} for Ursa Minor, Sculptor, Leo I, and Leo II; and \citet{coleman2008} for Fornax.}
\label{SF.fig} 
\end{figure*}

We find that three classical dwarf spheroids should have fallen in significantly later.  Leo I is the most recent at $\sim 2$ Gyr ago, followed by Leo II at $\sim 2-6$ Gyr and Fornax at $\sim 5-8$ Gyr look back time.  
Interestingly, all three of our late infall candidates demonstrate active star formation at later times than do our early-infall candidates. 
 Leo I \& II both have long epochs of star formation stretching from $\gtrsim 10$ Gyr ago to $\sim 2$ Gyr ago \citep{mighell1996,caputo1999,gallart1999,dolphin2002,bosler2007,orban2008,gullieuszik2009,lanfranchi2010}.  The end of star formation correlates well with the predicted infall time for Leo I and is consistent with the fairly broad infall range for Leo II.   Fornax had a major burst in star formation $3-4$ Gyr ago \citep{stetson1998,saviane2000,coleman2008,held2010,letarte2010}, which corresponds to a time just after our estimated infall.  

Both the LMC and SMC are still actively forming stars \citep{demarchi2011,harris2009} and are unique among the classical Milky Way satellites in containing neutral gas today \citep{grcevich2009,grcevich2010}.  According to our VL2 subhalo analysis, the high 3D speed and position of the LMC demand that its
accretion was more recent than $\sim 4$ Gyr.  Similarly, while the radial velocity and position of the SMC alone would seem to favor an early infall (see Fig. \ref{Tdists.fig}) its relatively high proper motion pushes us towards recent accretion $\lesssim 4-9$ Gyr albeit with large uncertainties.  Though we have not demanded that subhalos be interacting or massive in order to assocate them with the Clouds, it is at least encouraging that our analysis is consistent with the idea that they
were accreted fairly recently and at approximately the same time  \citep{besla2007,sales2011}.  

Fig. \ref{SF.fig} provides an illustration of the diversity of star-formation histories relative to the infall times in the classical dwarfs exemplified by a few specific cases (with names indicated in each panel).  
UMi, Sculptor, and Leo II  are examples of dwarfs that appear to be quenched just as they fell into the Milky Way halo.   While star formation in Sculptor (upper middle) goes out with a whisper, Leo I (lower middle) may or may not have experienced a burst of star formation after it fell in.  The remaining dwarfs, illustrated by Carina and Fornax in the right panels, seem to experience bursts of star formation immediately after infall.  In galaxy groups and clusters, many galaxies experience bursts of star formation triggered by the hot host halo compressing gas in the satellites \citep{gavazzi1985,bothun1986,bekki2003,mahajan2010,rose2010,abramson2011,santiago2011}.  Something similar may be occuring with the dwarf galaxies in the Milky Way halo.  

At this point, it is not clear exactly what terminates the star formation or what the time scale for termination is because the infall-time PDFs are fairly broad, but it is useful to speculate in order to define a starting point.   We suggest that the systems of the first category -- the ones that that seem to have star formation quenched just upon accretion -- are dominantly quenched by quick-acting ram pressure stripping \citep{gunn1972}.   The second class of objects, those with more gradual quenching after accretion, are more likely affected by strangulation \citep{larson1980}.  Finally, the bursting population may be so gas rich upon infall that they carry their material inward towards the Galactic center long enough to experience tidal shocking and associated bursting activity.  These scenarios are clearly simplified, as all of these processes should operate to some extent and it is not clear why one effect should dominate over another from case to case.  It may have to do with the details of the orbits or perhaps the arrangement of material within the infalling galaxy.   Nevertheless our infall time estimates provide an initial point of comparison.

\subsubsection{Ultrafaint dwarfs}
Are the patterns any different for the ultrafaint dwarfs?  One obvious difference between the ultrafaint and classical dwarf populations is that the ultrafaint dwarf stellar populations tend to be much older, with the exception of Leo T \citep{lee2003,irwin2007,dejong2008,dejong2008a,martin2008a,kuehn2008,okamoto2008,okamoto2008a,norris2008,greco2008,ryan-weber2008,lee2009,sand2009,sand2010,simon2010,aden2011,norris2010,lai2011}. Other potential outliers are Ursa Major II, Canis Venatici I and Leo IV which might have a small young ($\sim 2$ Gyr) component \citep{dejong2008a,sand2010}.  Nevertheless, if the correlation between the end of star formation and the infall time characteristic of the classical dwarfs (and their quenching mechanisms) persisted for the ultrafaint dwarfs, we would expect the dwarfs to have disproportionally large $t_{infall}$. 

Indeed, we find that a higher  percentage of the ultrafaint dwarfs we consider in this paper are {\em consistent} with an early infall (9/12; Sextans, Canes Venatici II Bo{\" o}tes I, Coma Berenices, Segue I, Ursa Major I, Ursa Major II, Willman I) compared to the classical dwarfs (5/9), where again we define early infall to be $t_{infall} > 8$ Gyr.
Yet the errors on the ultrafaint infall times are often large enough that intermediate-age infall times are allowed for many of them.  Only  three ultrafaint dwarfs (Ursa Major II, Bo{\" o}tes I, and Segue 1) have kinematics and positions that actually {\em demand} early infall.    

Possibly more interesting are the ultrafaint dwarfs that were likely accreted somewhat later.  Leo T is an outlier in the sense that it appears to be making its first passage into the Milky Way and it is the only Milky Way dwarf besides the Clouds to have a detection in neutral hydrogen \citep{ryan-weber2008,grcevich2009,grcevich2010}.  \citet{grcevich2009} find that Local Group dwarfs with galactocentric distances to either the Milky Way or M31 less than 270 kpc do not have detections in neutral hydrogen while those at greater distances have a neutral gass mass $> 10^5M_\odot$.  Leo T, at a Galactocentric distance of $\sim 410$ kpc,  is clearly a galaxy that has yet to be significantly influenced by the Milky Way.

There are interesting differences among the other later-infall ultrafaint dwarfs.  Canis Venatici I ($t_{infall} \sim 2-6$ Gyr) and Leo IV ($\sim 5-9$ Gyr) both have predominantly old stellar populations but both show evidence for star formation $\sim 2$ Gyr ago \citep{martin2008a,sand2010}.  Unless the infall time lies at the wings of the infall-time PDFs for these objects, these  galaxies were forming tiny numbers of stars ($2\%$ to 5\% of the total) after they fell into the Milky Way.  This is puzzling; a number of classical dwarfs continue to form significant numbers of stars up until $\sim 1$ Gyr after infall (except for Carina and the Clouds, which keep bursting), and are continuing to form stars from early times until quenching inside the Milky Way halo.  It is curious that Canis Venatici I and Leo IV both form a small number of stars after infall but that there was a long time gap between the formation of the ancient, dominant stellar population and the tiny young population.  Infall occurred sometime in this gap.  It means that although star formation was largely quenched before infall, these dwarfs held onto some gas that was prevented from forming molecular hydrogen until the dwarfs fell into the Milky Way halo.   Perhaps additional pressure from the Milky Way hot halo was necessary to trigger the formation of molecular hydrogen in those dwarfs.  In any case, the star-formation histories of Canis Venatici I and Leo IV are very much unlike the star-formation histories of the more recently accreted classical dwarfs.

Another interesting outlier is Hercules.  This dwarf galaxy shows no evidence for stars younger than 10 Gyr \citep{sand2009,aden2011}, yet its infall time is predicted to be $t_{infall} \sim 2 - 8$ Gyr.  This is the most significant offset we have found between probable infall time and the end of star formation other than the Clouds.  If our estimates of the infall time are accurate, it suggests that star-formation quenching \emph{must} occur before the galaxy falls into the Milky Way, which is different than what we find for the classical dwarfs.  This is different from all the other dwarf galaxies, for which we either cannot determine the order of star-formation quenching and infall or for which the end of star formation happens after infall.

This offset between the star-formation epoch and infall time of Hercules is even more interesting when one considers Leo IV.  Leo IV and Hercules have quite similar Galactocentric distances ($\sim 140-150$ kpc) and luminosities ($\sim 10^4L_\odot$), and both appear to have fallen into the Milky Way since $z=1$ (see Table 1).   Neither has any detectable neutral gas.  However, there is evidence for a small and young stellar population in Leo IV but Hercules is entirely ancient \citep{sand2009,sand2010}.  Why is this the case?  One possibility is because the central dark matter density of Hercules appears to be lower than all the other Milky Way
dwarfs \citep{strigari2008,aden2009}, which would place in among the smallest dark matter halos known to host a galaxy.    The difference in the star-formation history of these two otherwise similar systems may hint that Hercules represents some sort of transition in star formation or stochasticity in star formation in small objects.  Better infall-time PDFs might shed light on the origin of the differences in the stellar populations given the similarity otherwise of these two dwarf galaxies.   

\subsubsection{Trends and Future Discovery Potential}
So far, we have found several fundamental differences in the quenching of classical and ultrafaint dwarfs and one possible similarity.  In the cases in which we can distinguish the infall from star-formation epochs, the classical dwarfs appear to be quenched after infall but the ultrafaint dwarfs tend to be quenched for the most part long before infall (though uncertainties are large).  This may suggest that the morphology--density relation or star-formation--density relation do not exist in the same way for dwarf galaxies as they do for $L_*$ galaxies.  While quenching in the classical dwarfs appears to occur within the Milky Way halo (and hence induces a star-formation--density relation), for the ultrafaint dwarfs for which the infall time can be cleanly separated from the star-formation epoch, the quenching appears to precede infall.  This means that there is only a star-formation--density relation in that the reionization epoch is density-dependent, and various processes associated with reionization can quench star formation in small galaxies \citep{bullock2000,benson2002,busha2010}.  The one similarity is that it appears that a burst of star formation could be triggered in both classes of dwarfs after infall, but the degree to which that happens for both populations, and what that says about the specific quenching mechanisms after infall, are debatable.

An interesting question to ask is what types of dwarf galaxies we may expect to discover in next-generation deep wide-field surveys (e.g., LSST\footnote{http://www.lsst.org}), and what we may learn about star formation in these smallest of galaxies.  Since the sample of bright (classical) Milky Way dwarf galaxies is likely complete, the question really revolves around the population of Milky Way ultrafaint galaxies, and how representative the current sample is of the population as a whole. One of the issues with the currently known sample of Milky Way dwarf galaxies is that we currently can only find ultrafaint dwarf galaxies that are relatively nearby unless they are also relatively bright.  From Fig. \ref{RvsVr.fig}, we find that nearby galaxies disproportionately fell into the Milky Way early.   Thus, we expect that next-generation surveys, which will be complete to the Milky Way virial radius for dwarfs with the same surface brightness as already-discovered dwarfs, could find a number of faint dwarf galaxies that fell into the Milky Way more recently \citep{tollerud2008}.

There are already hints of interesting things in the current sample of dwarf galaxies, and it will be highly interesting to see if those hints become real trends when the catalog of Milky Way dwarfs is more complete.  Since most of the ultrafaints already discovered are close, and tend to have early infall times, it is difficult to tell whether star-formation quenching preceded or followed infall.  In other words, it is difficult to tell if the early-infall dwarfs are old because they fell into the Milky Way halo early and were then quenched, or if they were ``born old'' before they were accreted by the Milky Way.  It may be easier to tell the sequence of events for more recently accreted dwarfs since the separation of time scales may be larger.  If we find a number of more recent arrivals have only old stellar populations like Hercules, it will indicate either that feedback from those early epochs of star formation drove out all the gas in the galaxies or that the host halos were unable to accrete gas from the IGM.  This clearly did not happen with Leo T, but this is a fairly bright galaxy compared to something like Segue 1.  

If we discover new arrivals that are more similar to Canis Venatici I or Leo IV in that they have some young stars in addition to the predominantly old stellar population, we will learn that even the smallest galaxies may retain enough gas to late times to form some stars, even if formation of molecular hydrogen for star formation is suppressed.  It would be surprising but highly interesting if we found ultrafaint dwarf galaxies that were not dominated by old stellar populations, since all of the currently known ultrafaint dwarfs have overwhelmingly old stellar populations.  A number of classical dwarfs are dominated by intermediate-age populations (e.g., Fornax, Leo I, Leo II, Carina), and we should be able to learn at what mass-scale (either in stellar or dark-matter mass) these intermediate-age systems disappear if they do at all.  One worry to keep in mind is that the lowest mass dark matter halos will tend to host dwarfs that are the lowest surface brightness, and therefore the hardest to detect \citep{bullock2010}.    Nevertheless, if we see any diversity in the star-formation histories for recently accreted ultrafaint dwarf galaxies, it will tell us something about the stochasticity of star formation in small galaxies and possibly about the physical mechanism(s) thereof.

The questions we will likely be able to answer with a more complete catalog of Milky Way dwarf galaxies are the following:  Are old galaxies old because they were quenched when they fell into the Milky Way or because they were unable to accrete or hang onto their gas prior to infall?  How much is star formation quenched in small galaxies before they fall into the Milky Way?  What property of the satellites is most correlated with quenching before infall, and what will that tell us about the physical mechanism(s) for quenching before the satellites fall into the Galaxy?  How is star formation quenched once satellites are within the virial radius of the Milky Way?  Do the satellites at larger radius have later epochs of star formation due to their infall time or because quenching is more effective at smaller galactocentric distances?  How important is ram-pressure stripping for quenching star formation relative to other environmental factors?  Can we learn something about the Milky Way's hot halo of gas today or its evolution in the past?  How much does any of this depend on the mass of the dark-matter halos of the satellites?  

\begin{table*}
\label{accTimes.tab}
{\bf Table 1:} Observed and derived properties of the Milky Way dwarf satellite
galaxies considered in this paper.\\
\centering
\begin{tabular}{lccccccc}
Galaxy & Distance & Luminosity & Galactocentric Radius & $V_r$ or
$V_{\mathrm{los}}$(gsr) & $V_t$ & $V$ & $t_{infall}$\\
& [\kpc]   & [$\LsunV$] & [\kpc] & [\kps] & [\kps] & [\kps] & [\Gyr] \\
\hline \hline
Ursa Minor &  $77 \pm 4 \:^{(i)}$ & $3.9^{+1.7}_{-1.3} \times10^5\:^{(b)}$ &
$79 \pm 4$ & $-75 \pm 44\:^{(w)}$ & $144 \pm 50\:^{(w)}$ & $162 \pm 49$ &  $
8-11$ \\
Carina & $105 \pm 2\:^{(a)}$ & $4.3^{+1.1}_{-0.9} \times 10^5\:^{(b)}$ &
$107 \pm 2$ & $20 \pm 24\:^{(s)}$ & $85 \pm 39\:^{(s)}$ & $87 \pm 38$ & $
7-9$ \\
Sculptor &  $86 \pm 5\:^{(g)}$ & $2.5^{+0.9}_{-0.7} \times 10^6\:^{(b)}$&
$86 \pm 5$ & $79 \pm 6\:^{(v)}$ & $198 \pm 50\:^{(v)}$ & $213 \pm 46$ & $
7-9$\\
Draco  &  $76 \pm 5\:^{(c)}$ & $2.2^{+0.7}_{-0.6} \times 10^5\:^{(b)}$ &
$76 \pm 5 $ & $-97 \pm 4 $ & - & - & $\sim 8-10$\\
Sextans &  $96 \pm 3\:^{(h)}$ & $5.9^{+2.0}_{-1.4} \times 10^5\:^{(b)}$ &
$99 \pm 3$ & $72 \pm 6$ & - & - & $\sim 7-9$ \\
Fornax & $147 \pm 3\:^{(a)}$ & $1.7^{+0.5}_{-0.4} \times 10^7\:^{(b)}$ &
$149 \pm 3$ & $-31.8 \pm 1.7\:^{(t)}$ & $196 \pm 29\:^{(t)}$ & $199 \pm 27$ &
$\sim 5-9$ \\
Leo II & $233 \pm 15\:^{(e)}$ & $7.8^{+2.5}_{-1.9} \times 10^5\:^{(f)}$ &
$235 \pm 15$ & $22 \pm 4\:^{(u)}$ & $265 \pm 129\:^{(u)}$ & $266 \pm 129$ &
$\sim 1-6$ \\
Leo I  & $254 \pm 18\:^{(d)}$& $5.0^{+1.8}_{-1.3} \times 10^6\:^{(b)}$ &
$258 \pm 18$ & $177 \pm 5$ & - & - & $\sim 2$ 
\\ \hline
Ursa Major II &  $32 \pm 4\:^{(q)}$ & $4.0^{+2.5}_{-1.4} \times 10^3$ &
$38 \pm 4$ & $-17 \pm 3$ & - & - & $\sim 8-11$\\
Bo{\"o}tes I & $66 \pm 3\:^{(j)}$& $2.8^{+0.6}_{-0.4} \times 10^4$ &
$64 \pm 3$ & $107 \pm 2$ & - & - & $\sim 7-10$ \\
Segue 1 &  $23 \pm 2\:^{(m)}$ & $3.4^{+3.0}_{-1.6} \times 10^2$ &
$28 \pm 2$ &$111 \pm 4$ & - & - & $\sim 7-10$ \\
Ursa Major I &  $97 \pm 4\:^{(p)}$ & $1.4^{+0.4}_{-0.4} \times 10^4$ &
$102 \pm 4$ & $-11 \pm 3$ & - & - & $\sim 6-10$ \\
Coma Berenices &  $44 \pm 4\:^{(m)}$ & $3.7^{+2.2}_{-1.4} \times 10^3$ &
$45 \pm 4$ & $82 \pm 1$ & - & - & $\sim8-11$ \\
Leo IV & $160 \pm 15\:^{(m)}$& $8.7^{+5.4}_{-3.6} \times 10^3$ &
$161 \pm 15$ & $10 \pm 5$ & - & - & $\sim5-9$  \\
Canes Venatici I & $218 \pm 10\:^{(k)}$& $2.3^{+0.4}_{-0.3} \times 10^5$ &
$218 \pm 10$ & $78 \pm 2$ & - & - & $\sim 2-7$ \\
Hercules $^{(n)}$ & $133 \pm 6$& $1.1^{+0.5}_{-0.3} \times 10^4$ &
$127 \pm 6$ & $145 \pm 4$ & - & - & $\sim 2-8$ \\
Willman 1 &  $38 \pm 7\:^{(r)}$ & $1.0^{+0.9}_{-0.5} \times 10^3$ &
$43 \pm 7$ & $35 \pm 3$ & - & - & $\sim6-11$ \\
Canes Venatici II & $160 \pm 5\:^{(l)}$& $7.9^{+4.4}_{-3.0} \times 10^3$ &
$161 \pm 5$ & $-96 \pm 1$ & - & - & $\sim1-9$ \\
Leo T $^{(o)}$ & $407 \pm 38$& $1.4 \times 10^5$ & 
$412 \pm 38$ &  $-61 \pm 4$ & - & - & $< 1$ \\
\hline
SMC &  $61 \pm 4\:^{(z)}$ & $4.1 \times 10^8\:^{(z)}$  &
$58 \pm 4$ & $23 \pm 7\:^{(y)}$ & $301 \pm 52\:^{(y)}$ & $302 \pm 52$ & $\lesssim
4-9$\\
LMC &  $50 \pm 3\:^{(z)}$ & $1.4 \times 10^9\:^{(z)}$ & 
$49 \pm 3$ & $89 \pm 4\:^{(x)}$ & $367 \pm 18\:^{(x)}$ & $378 \pm 18$ & $\lesssim
4$\\

\end{tabular}
\vskip 0.5 cm Note: Galaxies are grouped from top to bottom as
pre-SDSS/classical MW dSphs followed by post-SDSS MW dSphs, with the
Magellanic Clouds at last. Columns 5-7 show radial, tangential and spacial
velocities in the galactic rest frame for those galaxies for which proper
motions are known. For those galaxies with unknown proper motions column 5 shows
the line of sight velocity in the Galactic Standard of
Rest (GSR) frame $V_{\mathrm{los}}$(gsr).\\

References: Except for Hercules and Leo T, values in column 3 (Luminosity) of
the post-SDSS MW dSphs are from \citet{martin2008b}. Values in columns 4
(Galactocentric Radius) and 5 ($V_{\mathrm{los}}$(gsr)) are derived and quoted
respectively from the NASA/IPAC Extragalactic Database.
The individual references are as follows: a) \citet{pietrzynski2009}, b) Derived
from apparent magnitudes listed in \citet{mateo98}, 
c) \citet{bonanos2004}, d) \citet{bellazzini2004}, e) \citet{bellazzini2005}, 
f) \citet{coleman2007}, g) \citet{pietrzynski2008}, h) \citet{lee2003}, i)
\citet{carrera2002}, j) \citet{dallOra2006}, k) \citet{martin2008a},
l) \citet{greco2008}, m) \citet{belokurov2007}, n) \citet{sand2009},
o) \citet{dejong2008}, p) \citet{okamoto2008}, q) \citet{zucker2006}, 
r) \citet{willman2005}, s) \citet{piatek2003}, t) \citet{piatek2007}, 
u) \citet{lepine2011}, v) \citet{piatek2006}, w) \citet{piatek2005}, 
x) \citet{kallivayalil2006a}, y) \citet{kallivayalil2006b}, z) NASA/IPAC
Extragalactic Database .
%\end{array}
%\end{table*}
\end{table*}

\subsection{Towards more robust and accurate infall-time PDFs}
Of course, the caveat to the discussion so far is that we have only examined the energy-infall relation and its consequences in one simulated dark-matter halo, and a halo simulated with $\sigma_8$ smaller by $\sim 0.07$ from the current preferred value.  Moreover, the actual mass of the Milky Way dark-matter halo is not known to better than a factor of two, so it is not clear if VL2 is even mass-wise (let alone mass-assembly-wise) a good match to the Milky Way \citep{wilkinson1999,battaglia2005,battaglia2006,dehnen2006,xue2008,reid2009,watkins2010}.  There is also some nontrivial uncertainty in the mass and distribution of stars in the Galaxy \citep{binney2008}.  The VL2 mass is on the high side, but certainly within range of what is expected for the Milky Way, given the existence of the Magellanic Clouds, which push expectations towards the $\sim 2 \times 10^{12}$ M$_\odot$ virial mass range \citep{boylan-kolchin2011}.

There are several things that are likely to matter for getting the energy-infall relation right for the Milky Way specifically.  The most important things may be both the dark and baryonic masses of the Milky Way, and the evolution of the baryonic component.  Both the dark and baryonic masses affect the gravitational potential and hence the energy normalization in the energy-infall relation.  In addition,  the normalization of the subhalo mass function and distribution in the halo depend on the dark and baryonic masses of the Galaxy \citep{donghia2010}.  The shape of the subhalo mass function depends on the baryons, too, since tidal shocking by the baryonic disk eliminates many of the lower-mass subhalos that otherwise would have been relatively unaffected by dynamical friction and tidal stripping \citep{donghia2010}.  The differences in the radial distribution and mass function of satellites may shift the infall-time PDFs at a given position in phase space.

Getting $\sigma_8$ might matter as well. Structure forms earlier for higher $\sigma_8$, and subhalos are accreted earlier for the currently accepted value of $\sigma_8$ rather than that used for VL2.  This might shift the slope of the energy-infall relation.  However, the gravitational potential of the host may also be different because of $\sigma_8$, and hence the typical subhalo energy at infall may also be different.  It will be important to see if the energy-infall relation depends on the underlying cosmology.

While it would be good to check that the mass-assembly history (for fixed halo mass) does not shift the energy-infall relation much, we suspect that this will not be the major driver for inflating the width of the infall-time PDFs.  This is because the Milky Way is likely to have had a relatively quiescent merger history for the majority of its life.  Simulations of disk galaxies have indicated that the thinness of the Milky Way stellar disk cannot be reproduced if the Milky Way had a cosmologically-typical 10:1 merger since $z=1$ \citep{purcell2009}.  As we showed in Sec. \ref{sec:einfall}, the gravitational potential in the outskirts of halos does not change much with time if halos are in the ``slow growth'' phase characteristic of quiescent halo evolution.  Since the halo potential does depend on the overall mass distribution of the dark-matter halo, we expect the halo mass rather than the mass-assembly history to drive the normalization and the slope of the energy-infall relation.

The main reason to simulate many Milky Way-type galaxies, even if there were no uncertainty in the Milky Way gravitational potential and its evolution, is to get better statistics on the subhalos.  Since VL2 has a relatively small number of halos, we had to choose relatively large windows in $r$ and $V_r$ to select subhalo samples for each Milky Way dwarf, and we only had samples of 20 subhalos for each dwarf (Sec. \ref{sec:accMW}).  If we had a much larger subhalo sample from which to work, we could select only subhalos with $r$ and $V_r$ within the observational uncertainties and have a large enough sample to create a smooth infall-time PDF.

On the observational side, the main way to improve the infall-time PDFs is to include proper-motion data.  As we have seen in Sec. \ref{sec:accMW}, proper-motion measurements sharpen the PDFs, but only if the uncertainty in the proper motion translates to $\sim 50\hbox{ km s}^{-1}$.  However, in the case of Fornax, there is still a double-peaked PDF in spite of an excellent proper motion measurement.  It is possible that better subhalo statistics will sharpen the PDF since the window in $r$ and $V_r$ from which we select a representative simulated subhalo sample for each dwarf will narrow.

The best thing to do, in terms of producing robust infall-time PDFs, is to simulate a number of Milky Way-type systems with different halo masses and with baryons.  This will give us both good subhalo statistics and an idea of how much the dark and baryonic masses affect the energy-infall relation since there are significant uncertainties in both for the Milky Way.  It would also be useful to get or improve proper-motion measurements of the observed dwarf galaxies.  

\subsection{Other consequences: stellar streams and dark matter flows}

We have focused on the energy-infall relation for surviving dwarf galaxies, but we may also apply this relation to their tidal debris and to those dwarfs that did not survive.  In particular, we expect there to be an energy-infall relation for tidal streams.  Unless dynamical friction or satellite-satellite encounters are important for a satellite, its energy should not change much after infall onto the Milky Way (as we showed in Sec. \ref{sec:einfall}).  This means that tidally stripped material should ``remember'' the satellite's infall energy.  

Since the shape of the tidal stream approximately marks out the orbit that the surviving part of the satellite should have had, one can estimate the energy of the progenitor satellite based on the shape of the tidal stream.  The next step would be to find infall-time PDFs for the progenitor satellite and compare that to the abundance pattern and inferred star-formation history of the stars in the streams.  Though more work needs to be done in order to check this possibility, it could prove to by particularly important for the efforts of near-field cosmology.  Specifically, the progenitor galaxies that formed the earliest, at times approaching the epoch of reoinization, are more likely to be accreted early and be shredded by now \citep[e.g.][]{bullock2005}.  If one of the goals of Milky Way studies is to use local observations to constrain high-redshift phenomena, it would be useful to be able to associate stellar streams with accretion (and ultimately, formation) times in as many ways as possible.

The energy-infall relation also explains the typical energy of the dwarf galaxy debris field discussed by \citet{lisanti2011}.  Those authors are concerned with characterizing the dark-matter particle population at the Sun's position in the Milky Way from the tidal debris of the subhalos surviving today that had identifiable progenitors in VL2 at $z=9$, which they define as the epoch of reionization.  They find that the this debris has a characteristic velocity at the Sun's position.  We would explain this characteristic velocity feature as arising from the energy-infall relation.  Since the Sun is deep within the Milky Way's potential well, we find from Fig. \ref{RvsVr.fig} that any surviving subhalos (and their associated tidal debris) should have been accreted at early times.  In other words, the infall-time PDF of surviving subhalos and their streams that might pass through the Sun's orbit is sharply peaked at high $t_{infall}$.  This means that the typical orbital energy today $\mathcal{E}$, and hence the typical dark-matter particle speed, is also peaked.

\section{Summary and Conclusions}\label{sec:conclusion}

%The goal of this work is to find a way to cleanly and simply determine the infall times of Milky Way satellite galaxies based on their present-day orbits.  The motivation for finding the infall times is to determine how star-formation quenching proceeds in these small galaxies, since the infall time marks the transition between quenching outside the Milky Way dark-matter halo and quenching due to the environment inside the Milky Way dark-matter halo.

We used a single simulation of a Milky-Way-like halo, the Via Lactea II (VL2) halo \citep{diemandetal2007,diemandetal2008,kuhlen2010}, to investigate the relationship between the kinematic properties of subhalos and their infall times.  We found that there exists a tight correlation between the binding energy of a subhalo to the host at $z=0$ and the time at which the subhalo last passed inward through the host virial radius (see Fig. 1).  We found that the origin of this ``energy-infall relation'' was the on-average near conservation of the energy of the subhalo from the time of infall.  This means that the orbits of the subhalos are determined by the mass-assembly history of the host, since the energy of the infalling subhalo must be similar to that of a particle on a circular orbit at the virial radius at the time of infall.

Assuming that the Milky Way has a similar mass-assembly history and gravitational potential as the VL2 halo, we were able to assign infall-time probability distribution functions for the known Milky Way satellite galaxies based on their present-day kinematics and the energy-infall relation.  We found reasonably peaked infall-time PDFs even in galaxies where proper motions were not available (and hence only had an upper limit on the binding energy from their radii and radial velocities), but the addition of proper motion constraints sharpen the PDFs considerably. 
For example, the orbital energies of Carina, Ursa Minor, and Sculptor are all strongly indicative of early accretion, more that 8 billion years ago.
Conversely,  Leo T, Leo I, and the LMC were all recently accreted, within the last few billion years.  Fornax, Leo II, and the ultrafaint dwarf Canes Venatici I are all examples of intermediate lookback time accretions.  Accretion time estimates for each dwarf are provided in Table 1.

When comparing the infall-time PDFs for individual Milky Way satellite galaxies with their inferred star-formation histories, we found a number of interesting trends.  For the classical dwarf galaxies, we found that the infall time occurred at a similar time as the end of star formation in about half the sample, but that star formation continued for a short while before ending in the other half.  Several of the dwarfs had at least one burst of star formation immediately after infall.  For the ultrafaint dwarf galaxies, we found that about half the galaxies fell in quite early and could not determine whether star-formation quenching occurred before or after infall.  Several of the dwarfs showed some evidence for a small amount of star formation after infall (Leo IV, Canis Venatici I, and possibly Ursa Major II), but Hercules clearly stopped forming stars long before it fell into the Milky Way.  In those cases, the bulk of star formation was quenched long before the dwarfs became satellites of the Milky Way.  The diversity of the offsets, both in magnitude and sign, in the infall time versus the end of star formation suggests that quenching is not a single and uniform process for these smallest of galaxies.  It means that any star-formation--density or morphology--density relation for the dwarf galaxies does not share the same origin as those relations for more massive galaxies.

The energy-infall relation that we have presented here will hopefully inspire explorations that benefit from a larger number of simulations, including eventually baryonic physics.  If the correlation between energy and infall time proves to be universal, even in some renormalized way, it will provide a new avenue for testing formation scenarios for satellite galaxies.  In this work we have taken the first tentative steps in that direction, part of the larger goal to put these interesting objects to use as the Rosetta Stones of galaxy formation that they have longed promised to be.

\section*{Acknowledgments}
MR was supported by a CONACYT doctoral Fellowship and NASA grant NNX09AG01G.  
A.H.G.P is supported by a Gary McCue Fellowship through the Center for Cosmology at UC Irvine and NASA Grant No. NNX09AD09G.  
We thank M. Kuhlen for providing particle data for the VL2 simulations along with J. Diemand and P. Madau for making the
VL2 data public. We thank Tuan Do for the discussions that initiated this work.

\bibliography{mybibs}

\end{document}